\documentclass[a4paper,fleqn]{cas-dc}

\usepackage{titlesec}
\usepackage{soul}

\titleformat{\section}
{\bfseries}
{\thesection.}{0.5em}{}

\titleformat{\subsection}
{\normalfont\itshape}
{\thesubsection.}{0.5em}{}

\titleformat{\subsubsection}
{\normalfont\itshape}
{\thesubsubsection.}{0.5em}{}

\usepackage[numbers]{natbib}
\usepackage{multirow}
\usepackage[]{algorithm2e}
\RestyleAlgo{ruled}
\usepackage{algpseudocode}
\usepackage{amsmath}
\usepackage{amssymb}
\usepackage{comment}

\usepackage{hyperref}
\hypersetup{
    colorlinks=true,
    citecolor = blue,
    linkcolor=blue,
    urlcolor=blue,
    anchorcolor = black
}
\usepackage{caption}
\usepackage{subcaption}
\usepackage{booktabs}
\usepackage{graphicx}

\usepackage[export]{adjustbox}

\def\tsc#1{\csdef{#1}{\textsc{\lowercase{#1}}\xspace}}
\tsc{WGM}
\tsc{QE}
\tsc{EP}
\tsc{PMS}
\tsc{BEC}
\tsc{DE}

 \newlength{\saveheight}             
\begin{document}
\let\WriteBookmarks\relax
\def\floatpagepagefraction{1}
\def\textpagefraction{.001}

\shorttitle{ }

\shortauthors{\c{C} Demirel et~al.}

\title [mode = title]{Single-channel EOG-based human-machine interface with exploratory assessments using harmonic source separation}         





\author[1,2]{\c{C}a\u{g}atay Demirel}

\cormark[1]

\credit{Conceptualization of this study, Methodology, Interface, Data curation, Formal Analysis, Writing - original draft, Writing - review and editing}


\address[1]{Computer Engineering Department, Istanbul Technical University, Maslak, 34467 Sarıyer, Istanbul, Turkey}

\author[2]{Livia Regu\c{s}}

\credit{Visualization, Writing - review and editing, Formal Analysis}

\author[1,3]{Hatice K\"{o}se}

\credit{Supervision, Project administration}


\address[2]{Donders Institute for Brain, Cognition and Behaviour, Kapittelweg 29, Nijmegen, 6525 EN, Netherlands}

    
\address[3]{AI and Data Engineering, Istanbul Technical University, Maslak, 34467 Sarıyer, Istanbul, Turkey}


\nonumnote{$^{*}$Corresponding author}

\nonumnote{\textit{E-mail address: }\href{mailto:cagatay.demirel@donders.ru.nl}{cagatay.demirel@donders.ru.nl} (\c{C}. Demirel).}
\let\printorcid\relax 

\begin{abstract}
There have been many studies on intelligent robotic systems for patients with motor impairments, where different sensor types and different human-machine interface (HMI) methods have been developed. However, these studies fail to achieve complex activity detection at the minimum sensing level. In this paper, exploratory approaches are adopted to investigate ocular activity dynamics and complex activity estimation using a single-channel EOG device. First, the stationarity of ocular activities during a static motion is investigated and some activities are found to be non-stationary. Further, no statistical difference is found between the envelope sequences in the temporal domain. However, when utilized as an alternative to a low-pass filter, high-frequency harmonic components in the frequency domain are found to improve contrasting ocular activities and the performance of the EOG-HMI-based activity detection system substantially. The activities are trained with different classifiers and their prediction success is evaluated with leave-one-session-out cross-validation. Accordingly, the two-dimensional CNN model achieved the highest performance with the accuracy of 72.35\%. Furthermore, the clustering performance is assessed using unsupervised learning and the results are evaluated in terms of how well the feature sets are grouped. The system is further tested in real-time with the graphical user interface and the scores and survey data of the subjects are used to verify the effectiveness.

\end{abstract}




\maketitle

\section{Introduction}
\vspace{3mm}

Due to an aging population and the increase in chronic health problems worldwide, global reports on disabilities has raised concerns. Attempts have been made to overcome these motor disabilities through intelligent robotic systems that help people \citep{b1}. Moreover, there are various technological advancements in developing interactive assistive devices, which have accelerated in recent years. These include automated wheelchairs, verbal and postural controlled mechanics \citep[]{b2,b3,b4}. In contrast, people with locked-in syndrome (LiS) - a rare neurological illness resulting in partial or complete stroke \citep{b5}, due to injured parts of the brainstem. Additionally, some patients with motor degenerative diseases, namely amyotrophic lateral sclerosis (ALS) or multiple sclerosis (MS), are unable to operate these devices \citep{b6}. As a consequence, those suffering from such diseases require non-invasive human-machine (HMI) based assistance as an alternative way for regaining their mobility \citep{b7}.

The majority of the HMI systems consist of two possible inputs: electroencephalogram (EEG) or electrooculogram (EOG). There are numerous benefits of using EOG over EEG based HMI pipelines including significantly less expensive and lightweight equipment, higher temporal resolution, and smaller data streaming with higher discriminant factor over detection of an ocular event. It has been shown that single-channel EOG devices could achieve the same goal of bulky EEG options to yield a variety of HMI systems \citep[]{b7, b8}. Higher precision activity around the ocular muscles is clearer information to represent an input action. Moreover, it will decrease the need of artifact reduction \citep{b9}, given that EOG activity itself could be considered as an artifact voluntarily created by the user. 

Even though EOG setup requires relatively less effort compared to EEG, still many of the systems use wet or saline based electrodes with multiple channels for better referencing across eyes \citep{b10}. Time consuming and cumbersome set up, circumvents the goal of an effortless solution for impaired patients. Given that they have limited capabilities and skills for device placement, single-channel EOG based HMI approaches would be a logical option to follow \citep{b11,b12}.

Further, the majority of studies developing EOG-HMI based systems have developed decision making models based on blink and saccadic potentials \citep{b12.1}. As a result of these findings relying on EOG time-domain potential oriented rule-based systems, complex ocular activity recognition haven't been developed using single-channel EOG signal.

Accordingly, some multi-channel designs have been developed that are not practical but can provide acceptable real-time performance. Using Ag-AgCl surface EMG electrodes, four different activities (right, left, up, down) are estimated from a five-channel system \citep{b12.2}. In another study, five different eye movements are estimated using 6 Ag/AgCl EMG electrodes \citep{b12.3}. There are also some studies that estimate only eye blinks with a single-channel EOG \citep{b12.4}\citep{b12.5}. The conclusion that can be drawn from all these studies is that minimal-sensing (single-channel) EOG devices are avoided to provide comprehensive control.These either employ saline-based multiple EMG surface electrodes or rule-based control is provided to use simple eye-blink artifacts.

In the proposed system, a single-channel EOG headset is used to create the HMI system with additional exploratory assessments of various ocular activities (normal glance, left eye closed, right eye closed, frowning, eyebrows up, blinking) using the single-channel EOG device. The similarities and differences between the various ocular activities are then explored using statistical, supervised, and unsupervised learning methods. Accordingly, the graphical user interface (GUI) is created to control a mouse cursor in real-time.

\section{Experiments \& Methods}\label{ch:ifnecch5}

\subsection{Overview of the EOG-HMI system}
\vspace{3mm}

\begin{figure}[htbp]
 \centering
 \includegraphics[width=0.6\columnwidth,keepaspectratio=true]{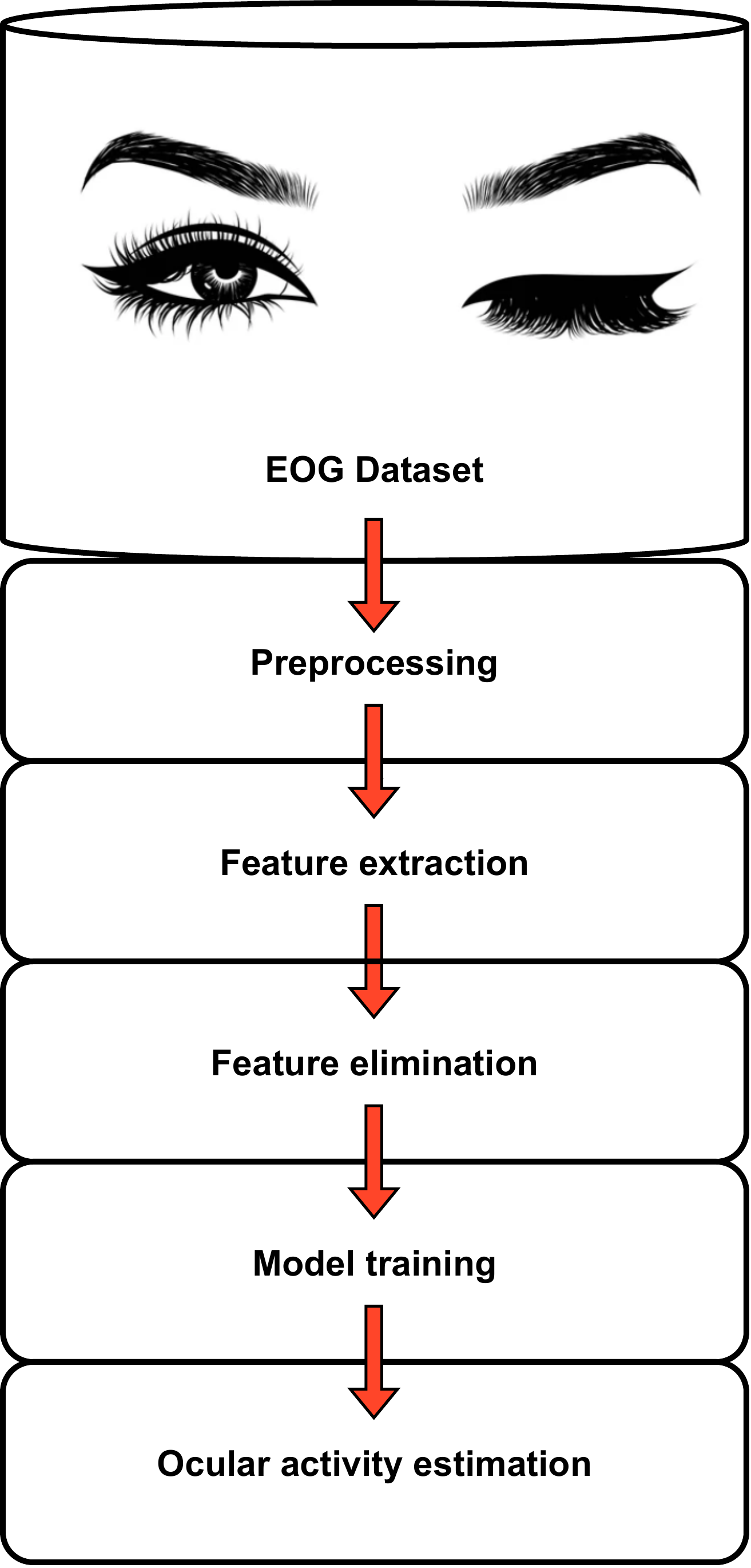}
 \caption{Overall flow diagram of EOG-HMI pipeline}
  \label{fig:overallFlowDiagram}
\end{figure}

Each subject is initially instructed to carry out all visual functions individually.
The preprocessed collected EOG signals undergo feature extraction, feature elimination, and low discriminative feature removal. The feature set is trained using several machine learning models and placed to the test in the final step. Overall flow diagram of EOG-HMI pipeline is shown in \autoref{fig:overallFlowDiagram}.

\subsection{Dataset Collection}
\vspace{3mm}

\begin{figure*}
 \centering
 \includegraphics[width=0.8\textwidth,keepaspectratio=true]{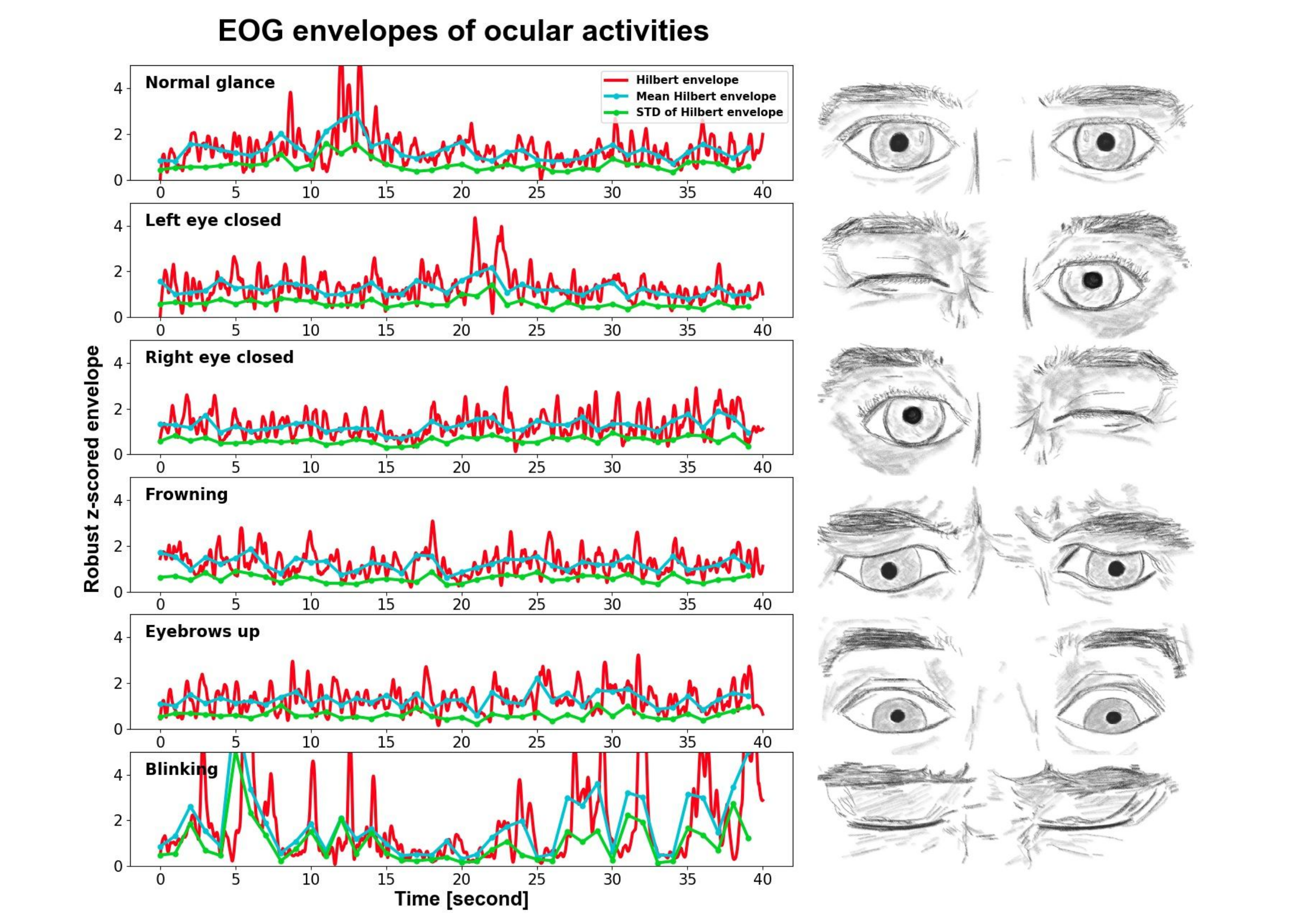}
 \caption{Ocular activities with corresponding EOG envelope signals. The left panel displays the robust z-scored envelopes of the ocular activities: Hilbert envelope, mean Hilbert envelope, and the standard deviation (STD) of the Hilbert envelope. To the right are displayed the corresponding ocular actions: normal glance, left eye closed, right eye closed, frowning, eyebrows up and blinking (from top to bottom).}
  \label{fig:eyeActions}
\end{figure*}

EOG data are collected using the NeuroSky MindWave Mobile Headset (NeuroSky, San Jose, CA, USA) headset on six participants, including four men and two women. The subjects are between the ages of 26 and 28, and five of them are right-handed. Each subject is provided six different eye actions, including: left eye closed, right eye closed, frowning face, instant eye blinks, eyebrows up, and normal glance. The eye actions are shown visually in ~\autoref{fig:eyeActions}. All procedures performed in this study involving human participants are in accordance with the Istanbul Technical University (ITU) ethics committee and with the 1964 Helsinki Declaration and its later amendments or comparable ethical standards.

Each ocular activity lasted 60 seconds in total, including a 20-second practice period and a 40-second actual experimental period, during which the subjects practiced the activities continuously without rest. While 5 of the activities are static (except blink), the subjects are asked to perform the ocular activities by fixing their gaze in the center of the screen without moving their eyes. In blink, subjects performed free eye blinking, again at dynamic intervals for 60 seconds. Each subject's EOG data are acquired for a total of 6 sessions, representing 240 seconds of experiment time. Sampling rate (Fs) of the EOG device is 500 Hz and 16 bit depth.

The recorded raw data are divided into 400 ms windows with padding of 50 ms. Each window is labeled with the corresponding eye action and the data set is created.

\subsection{Harmonic/Percussive source separation}
\vspace{3mm}

\begin{figure*}
 \centering
 \includegraphics[width=1\textwidth,keepaspectratio=true]{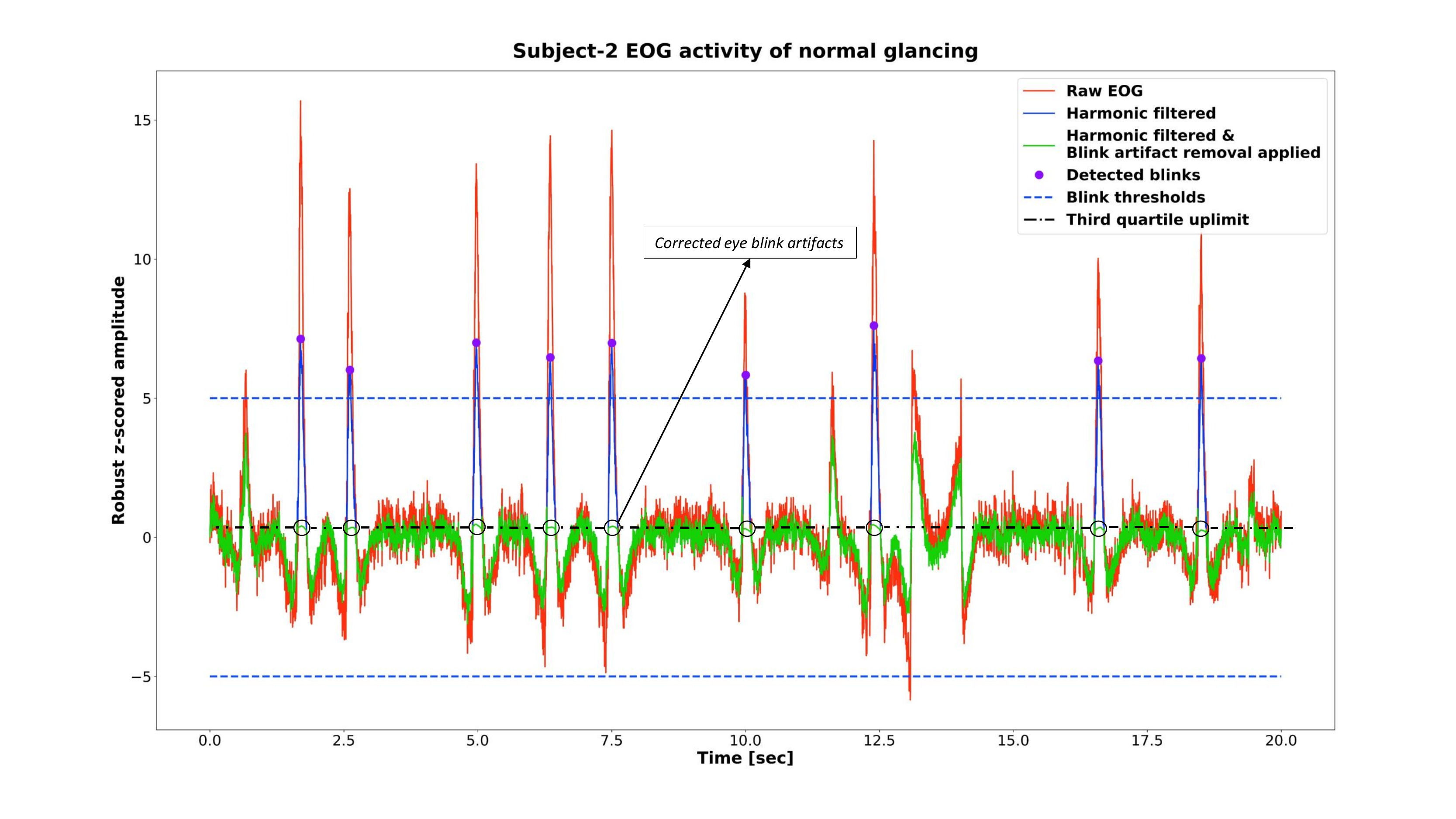}
 \caption{Comparison of harmonic filter and blink aftifact correction algorithm on time-domain signal. After a preliminary preprocessing stage, the signal is cleaned of percussive elements by removing blink artifacts with the HPSS filter. All points above the peak and the top of the third quartile of the narrow fraction surroundings  are replaced with the same indices in the envelope of the same signal using the novel blink artifact correction technique.}
  \label{fig:originalFilteredSignal}
\end{figure*}

Initially, 50 Hz notch filter with 2 Hz high-pass filters are applied to raw EOG signals to hinder the drift artifacts. These are artifacts caused by sweat, which are irrelevant for ocular or neural activities \citep{b13}. However, low-pass is not implemented to be able to operate high-frequency components reflected from fundamental frequencies (F0), related to ocular activity. The peaks that are an exact multiple of each other in the spectral domain constitute the harmonical representation of a given activity. In this regard, a harmonic/percussive source separation (HPSS) algorithm is applied to hinder percussive components from the spectral domain of the signal via short-time fourier transform (STFT) \citep{b13.1}. Those components do not follow the harmonic pattern and act as an environmental noise of time-series. This procedure is mostly used in audio science to decompose audio mixture. Utilizing the anisotropic smoothness of power spectrograms or the fact that harmonic power spectrograms are continuous in the time direction and percussion power spectrograms are continuous in the frequency direction, is one of the key techniques to HPSS \citep{b14}. Harmonic events are highlighted in a given EOG signal by removing continuous noisy sequence in frequency domain associated as percussive components. Moreover, the design of the algorithm, which is based on a median filter, also plays a role in eliminating signal drift. After using HPSS, the time-domain signal is obtained by reconstructing the spectral domain signal using an inverted short-time fourier transform (ISTFT). Formula of the harmonic filter is shown in Eq.~(\ref{eq:harmonicFilter}). Examples of raw, percussive and harmonic filtered EOG signals are shown in \autoref{fig:originalFilteredSignal}.

\begin{equation}
\centering
\begin{gathered}
H_i = M(S_h,l_{(harm)})\\
\label{eq:harmonicFilter}
\end{gathered}
\end{equation}

\noindent
where $M$ is a median filter, $S_h$ is a median filtering frequency piece, $l_{(harm)}$ is the length of the harmonic median filter, and $H_i$ is the harmonic filtered signal in frequency spectrum. Median filters transform the given signal piece into the median of the signal values. The median filtering formula is shown in Eq.~(\ref{eq:medianFilter}). 

\begin{equation}
y(n) = median(x(n-k:n+k), k=(l-1)/2)
\label{eq:medianFilter}
\end{equation}

The input vector is given as $x(n)$ and the output $y(n)$ is a median filtered signal of length $l$. Here $l$ represents the number of samples with respect to which the median filtering algorithm is implemented and $l$ is an odd number.

\begin{figure}[htbp]
 \centering
 \includegraphics[width=0.8\columnwidth,keepaspectratio=true]{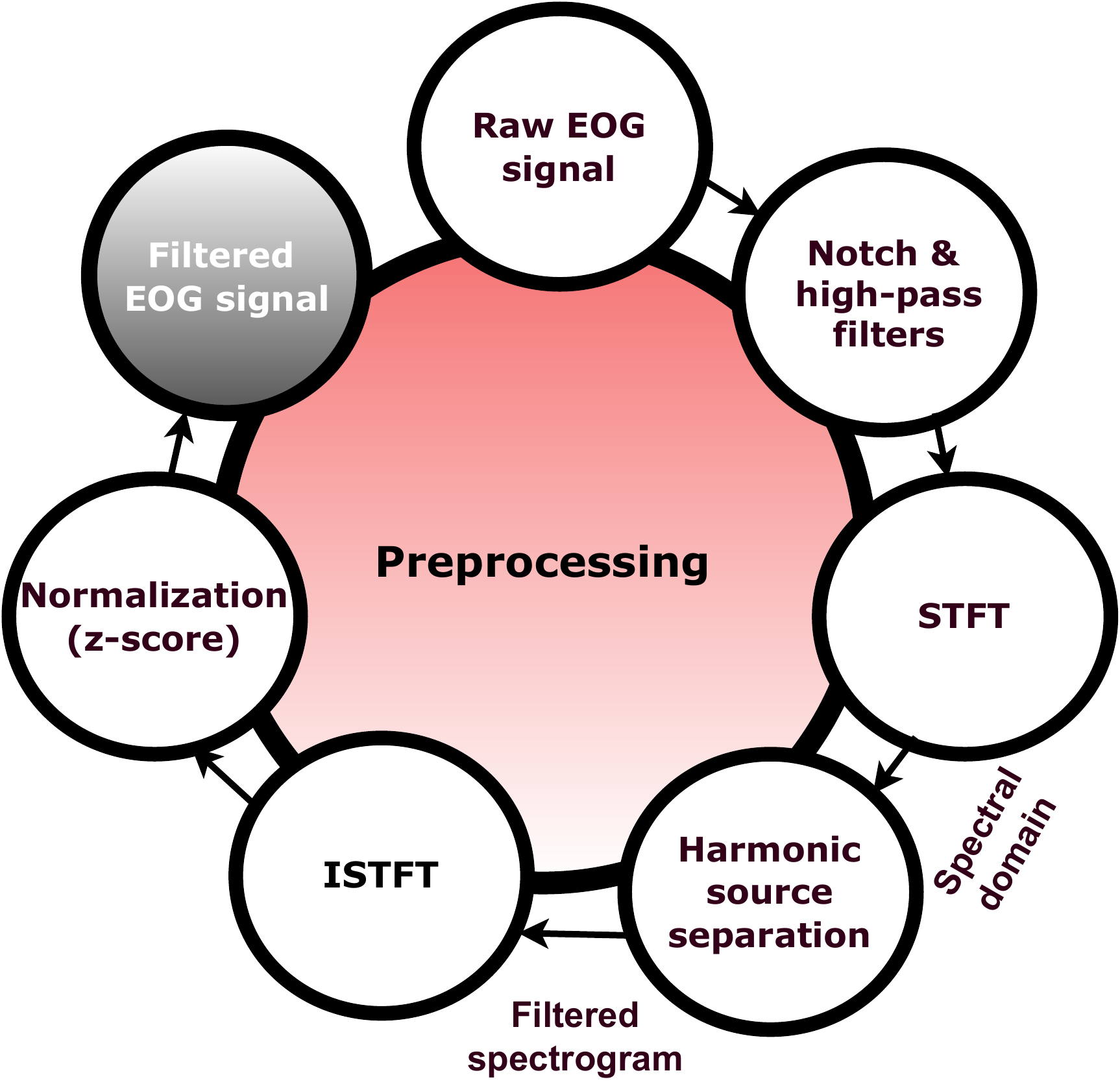}
 \caption{Noise Reduction Flow Diagram}
  \label{fig:NoiseReductionFlowDiagram}
\end{figure}

The flow diagram of the noise reduction process is shown in \autoref{fig:NoiseReductionFlowDiagram}.

\SetKwInput{KwProcess}{Process}

\subsection{Envelope insertion based blink artifact correction algorithm}
\vspace{3mm}

Blinking is a biological process - the rapid closure of the eyelid in a semi-autonomous manner \cite{b14.1}. It occurs spontaneously and may result from volitional commands, or as a reflex response to appropriate stimuli. Spontaneous blinking is done without external stimuli and is performed in the premotor brainstem without conscious effort. However, another blink passes as the reflex blink and is triggered depending on external stimuli. Just like the spontaneous blink, the reflex blink represent an unconscious eye movement and occurs faster than the spontaneous blink. However, in a comparative study, no statistical difference was observed between spontaneous and corneal reflexive blinks \citep{b14.2}. In another study, spontaneous blinks and voluntary blink activities were compared, and it was observed that the mean down phase range of voluntary blink activity was significantly higher \citep{b14.3}.

Based on all these past studies, it was considered that spontaneous and reflex blinks could not be prevented, and because they were different from voluntary eye movements. However, the end-goal HMI system is expected to be able to distinguish voluntary blinks from reflex or spontaneous blinks. Thus, the subjects blinked occasionally during the experiments, generating blinking artifacts. The EOG data obtained from five different eye movements (left eye closed, right eye closed, frowning, muscles up and normal gaze) contain blink artifacts as well. However, the end-goal HMI system is expected to be able to distinguish voluntary blinks from reflex or spontaneous blinks.

After the eye movement data received are passed through basic preprocessing and harmonic filtering, the third step is to correct the artifacts due to involuntary blinking. Since each EOG data are subjected to robust z-score normalization, the amplitude levels are almost at the same level for all subjects and all ocular activities. However, as a result of visual inspection, a lower limit is determined to detect blink artifacts. Peaks above this limit are considered as blink artifact if they are at a certain distance from each other. Eye artifacts are determined automatically and envelope insertion based blink artifact correction algorithm - a novel method, is designed. As seen in \autoref{alg:recursive_blink_artifact_correction_algorithm}, positive and negative peaks ($p\_ind$, $p\_ind\_neg$) are determined based on the lower ($q_1$) and upper quartile ($q_3$) limits of the whole signal. The area within these two boundaries is defined as the interquartile range and determines the inlier region of the EOG data during a repetitive task. In the subsequent step, the decimal logarithm of the signal is taken; the artifacts are presented, and the entire signal is enveloped with a Savitsky-Golay filter.

Next, the fraction of narrow band parts on the right and left sides of the determined peak are taken between the $q_3$ limit in the positive direction and the $q_1$ limit in the negative direction. This region is determined as the area where the artifact is effective. Each detected artifact region is replaced with the enveloped part of the corresponding region within a loop. In this way, artifact gaps in the EOG time-series are reconstructed without losing the semantic meaning of the activity-related signal.

\begin{algorithm}[htbp]
\caption{Envelope insertion based blink artifact correction algorithm}
\label{alg:recursive_blink_artifact_correction_algorithm}
\KwData{\\$ \;\;\;\;\;\;\; p_{ind} = find\_peaks(signal)$ \\
$\;\;\;\;\;\;\; p_{ind\_neg} = find\_peaks(signal * -1)$ \\
$\;\;\;\;\;\;\; q_3 = 3rd\_percentile$ \\
$\;\;\;\;\;\;\; q_1 = 1st\_percentile$}

\vspace{2 mm}

\KwResult{\\$ \;\;\;\;\;\;\; signal$}

\vspace{2 mm}

\KwProcess{\\$signal = \log_{10} {abs(signal)} $ \\
$signal\_envelope = sav\_gol(signal) $ \\

\vspace{2 mm}

\For{$i \;\; in \;\; length(peak\_indices)$}
 {
  $b\_d = 0$\\
  $e\_d = 0$\\
  
  \vspace{2 mm}
  
  \While{$signal[p_{ind}[i] - b\_d] > q3$}
   {
    $b\_d = b\_d + 1$
   }
   
   \vspace{2 mm}
   
  \While{$signal[p_{ind}[i] + e\_d] > q3$}
   {
    $e\_d = end\_d + 1$
   }
   
   \vspace{2 mm}
   
  $b\_i = p_{ind}[i] - b\_d$\\
  $e\_i = p_{ind}[i] + e\_d$\\
  $signal[b\_i : e\_i] = signal\_envelope[b\_i : e\_i]$\\
 }

\vspace{2 mm}

\For{$i \;\; in \;\; length(p_{ind\_neg})$}
 {
  $b\_d = 0$\\
  $e\_d = 0$\\
  
  \vspace{2 mm}
  
  \While{$signal[p_{ind\_neg}[i] - b\_d] < q1$}
   {
    $b\_d = b\_d + 1$
   }
   
   \vspace{2 mm}
   
  \While{$signal[p_{ind\_neg}[i] + e\_d] < q1$}
   {
    $e\_d = e\_d + 1$
   }
   
   \vspace{2 mm}
   
  $b\_i = p_{ind}[i] - b\_d$\\
  $e\_i = p_{ind}[i] + e\_d$\\
  $signal[b\_i : e\_i] = signal\_envelope[b\_i : e\_i]$\\
 }
 
\vspace{2 mm}

\Return \textit{(signal)}

}
\end{algorithm}

\subsection{Feature Extraction}
\vspace{3mm}
The preprocessed continuous EOG signals are divided into one-second windows with 500 ms padding. A 29-dimensional feature set that characterizes the temporal and spectral features of each window is extracted for all EOG acquisitions corresponding to the ocular activities. The features used in this study are shown below:

\begin{itemize}
\item Zero-Crossing Rate (ZCR) \citep{b15} : ZCR is the rate of sign-changes along signal. ZCR formula is shown in Eq.~(\ref{eq:zcr}).\\

\begin{equation}
ZCR = \frac{1}{T-1}\sum_{t=1}^{T-1}1_{R<0}(S_t, S_{t-1})
\label{eq:zcr}
\end{equation}

where $S_t$ is a signal point at time $t$ and the signal has a length of $T$. $R<0$ is an indicator function to only select the signal's time points changing their amplitude from positive to negative.

\vskip 2mm
\item Short-Term Energy \cite{b15}: it is the sum of squares of the samples in the given window, with its formula shown in Eq.~(\ref{eq:ste}).\\

\begin{equation}
E_T = \frac{\sum_{m=0}^{N-1}s^2(m)}{N}
\label{eq:ste}
\end{equation}

where $N$ is the total size of the discrete window and $s(m)$ is the energy of the given time point at $m$.

\vskip 2mm
\item Short-Term Energy Entropy  \citep{b15}: it is the entropy of averaged sub-frame energies, shown in Eq.~(\ref{eq:stee}).\\

\begin{equation}
\begin{split}
\label{eq:stee}
E_{tot} = \sum_{m=0}^{N}S(m)^2 \\ 
E_{sub} = \sum_{b=0}^{N\div K}\sum_{a=0}^{K}S(a)^2 \\ 
E_{entropy} = -\sum \frac{\sum E_{sub}^2}{E_{tot}} \log(\frac{\sum E_{sub}^2}{E_{tot}})
\end{split}
\end{equation}

\vskip 2mm
\item Spectral Entropy \citep{b15}: Formula of Spectral Entrophy is shown in Eq.~(\ref{eq:se}).\\

\begin{equation}
\begin{split}
\label{eq:se}
x_i = \frac{X_i}{\sum_{i=1}^N X_i} for \, i= 1 \, to \, N \\
E_{specentrophy} = - \sum_{X} x_i  \log_{2}x_i 
\end{split}
\end{equation}

\vskip 2mm
\item Spectral Centroid \citep{b16}: it is a metric for characterizing a spectrum. It indicates the location of the center of mass of the spectrum. The formula of the Spectral Centroid is shown in Eq.~(\ref{eq:scentroid}).\\.

\begin{equation}
C = \frac{\sum_{n}^{N-1}f(n)x(n)}{\sum_{n=0}^{N-1}x(n)}
\label{eq:scentroid}
\end{equation}

where $x(n)$ represents the weighted frequency value and $f(n)$ represents the center frequency of a bin.

\vskip 2mm
\item Spectral Bandwidth \citep{b17}:Formula of p'th order Spectral Bandwidth (SB) is shown in Eq.~(\ref{eq:sb}).\\

\begin{equation}
SB = \sum_{K} S(k) \, (f(k) - centroid^p)^{\frac{1}{p}}
\label{eq:sb}
\end{equation}

where $S(k)$ represents the energy bin of the spectrum, $f(k)$ represents the center frequency of that bin and $p$ is the order of spectral bandwidth.

\vskip 2mm
\item Spectral Roll-Off \citep{b18}: the roll-off frequency is defined as the central frequency of a spectrogram bin that contains at least 0.85 (by default) of the frame's energy. 

\vskip 2mm
\item Spectral Contrast \citep{b15}: SC feature estimates the strength of spectral peaks, valleys, and their differences in each frequency bin of the spectrogram. The formula of SC is shown in Eq.~(\ref{eq:scontrast}).\\

\begin{equation}
\begin{split}
\label{eq:scontrast}
Peak_k = \log{\frac{1}{aN}\sum_{i=1}^{aN}x_{k,i}} \\
Valley_k = \log{\frac{1}{aN} \sum_{i=1}^{aN}x_{k,N-i+1}} \\
SC_{k} = Peak_k - Valley_k
\end{split}
\end{equation}

where $N$ represents total number in k-th frequency bin, $a$ is set to be 0.02 as default, $x_{k,i}$ is the energy value of the frequency in sub-band $k$ as descended order.
\vskip 2mm
\item Polynomial Features \citep{b20}: it mainly provides k-th order polynomial fitting of the frequency spectrum of a given signal. Coefficients of found polynomials are accepted as extracted features.

\vskip 2mm
\item Maximum peak amplitude value (PAV) \citep{b20.1}: it measures the peak signal amplitude.
\item Maximum valley amplitude vaue (VAV) \citep{b20.1}: it measures the lowest negative amplitude value.
\item Areas under curve (AUC) \citep{b20.1}: it is the total energy of positive and negative absolute amplitude values.
\item {Statistical features}
\begin{itemize}
\item {Kurtosis}
\item {Skewness}
\item {Mean}
\item {Median}
\item {STD}
\end{itemize}

\item Coefficient of variation \citep{b21}: it is a statistical metric that is frequently used to calculate diversity of a series by taking the ratio of the STD to mean.

\item Wavelet decomposition \citep{b22}: The examination of localized fluctuations in power within a time series is being done using wavelet analysis. It can be defined as a discrete wavelet transformation (DWT) to reconstruct the waveform via convolving with low and high-pass filters. Instead of taking the whole filtered signal as a feature, statistical time-serie representations (mean, STD, some of the squared amplitudes, and entropy value) are extracted from the transformed signal as an additional feature space.

\vskip 2mm

\end{itemize}

\subsection{Normalization}
\vspace{1mm}

The feature set is z-score normalized with the normalization shown in Eq.~(\ref{eq:z-scoreNorm}).

\begin{equation}
Normalized(s_{i}) = \frac{s_{i} - \mu(S_{i})}
{\sigma({S_i})}
\label{eq:z-scoreNorm}
\end{equation}

\noindent
where $S_i$ is feature sequence of $features_i$ and $s_i$ is one sample that will be normalized.

\section{Analysis and Results}

\subsection{Stationarity assessment}
\vspace{3mm}

Stationarity in a time series is defined as the stability of the attributes defining the series over time and the absence of positive or negative correlations with time \citep{b24}. This is especially true if there is a continuous activity process that generates the time series, and any fraction of data in that process should not contain attributes that are time-dependent. If this is not the case and a series follows a particular trend over time, this indicates that the activity is dynamically changing along with the time. 

In the present study, all ocular activities are static except for the ones involving keeping one of the eyes closed. They should not have a unit root if linearity in the signal flow is maintained. In this case, it would mean that EOG activity has time-invariant and time-uncorrelated features at almost all time intervals. 

\begin{figure*}
     \centering
     \resizebox{\textwidth}{!}{%
     \begin{subfigure}[t]{0.9\textwidth}
         \centering
         \includegraphics[width=0.9\textwidth]{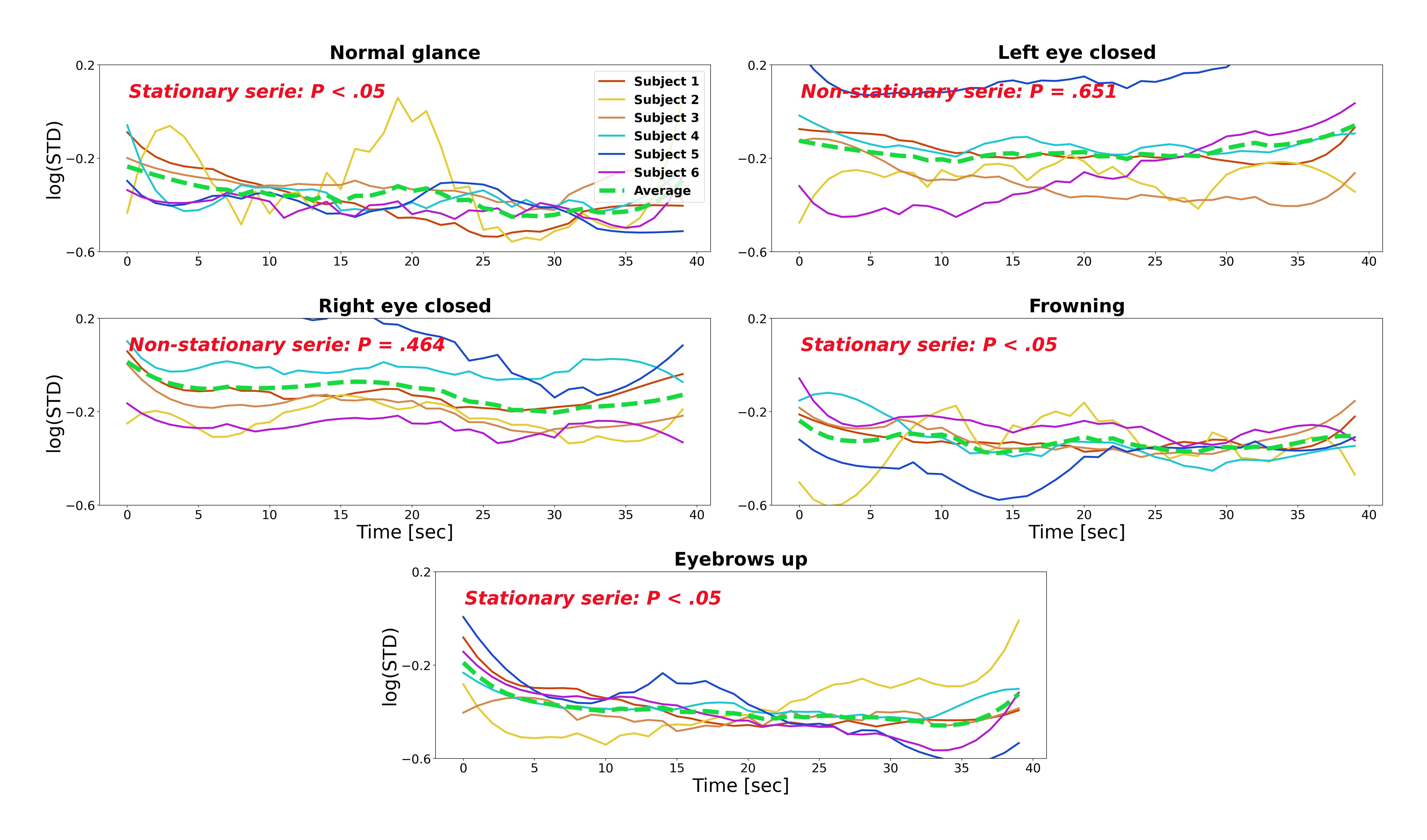}
         \caption{STD of Hilbert envelopes.}
     \end{subfigure}%
     }
     \newline
     \resizebox{\textwidth}{!}{%
     \begin{subfigure}[t]{0.9\textwidth}
         \centering
         \includegraphics[width=0.9\textwidth]{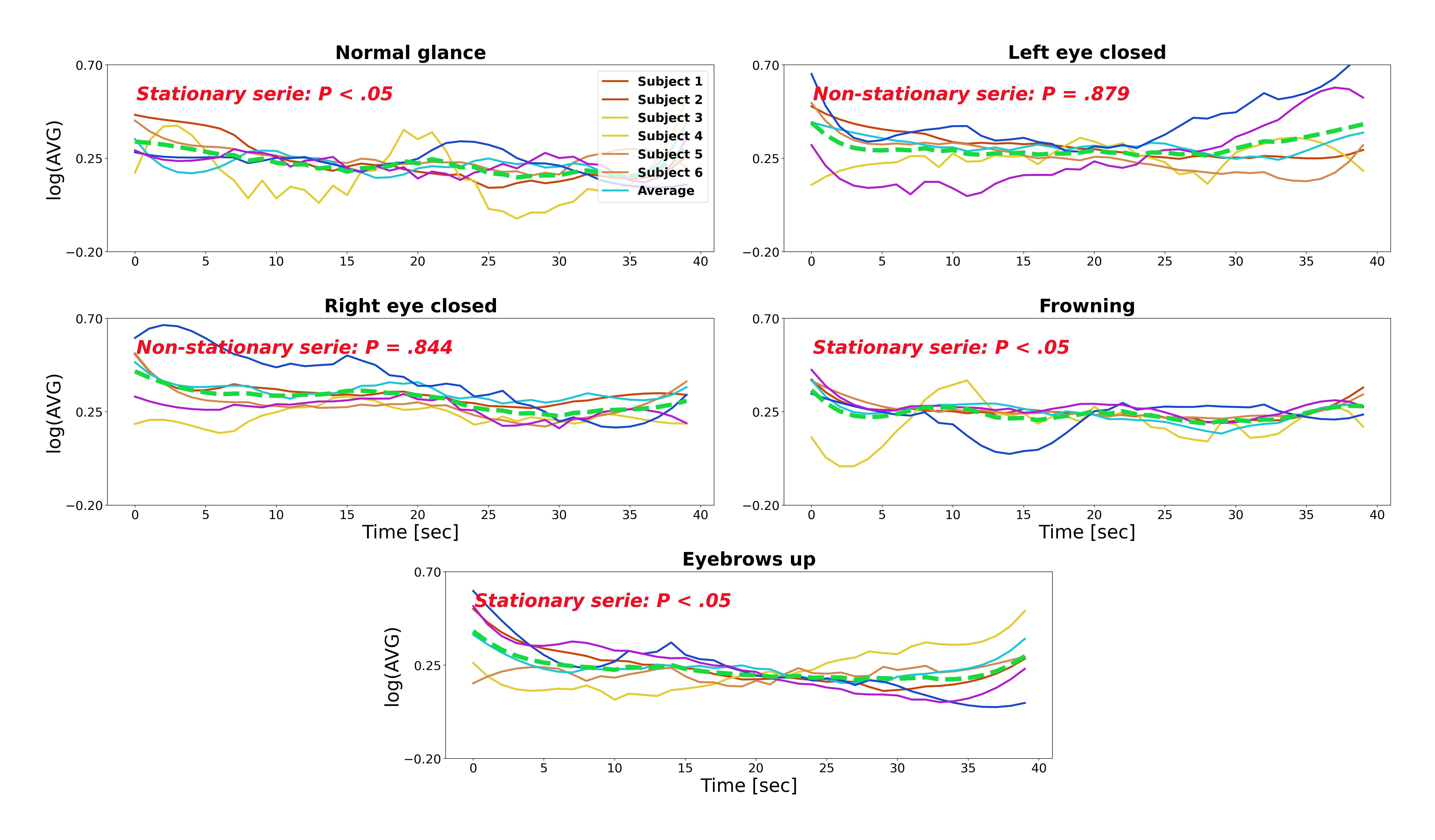}
        \caption{AVG of Hilbert envelopes.}
     \end{subfigure}%
     }
\caption{Stationarity check of static ocular activity time-series via temporal features. STD and AVG of epoched with logaritmized Hilbert envelopes are shown in \textbf{(a)}, and \textbf{(b)} respectively. Contrary to expectations, the two static ocular activities (left eye closed and right eye closed) are non-stationary series with insignificant results in the ADF test for both AVG and STD sequences.}
\label{fig:stdavg_envelopes_among_ocularactivities}
\end{figure*}

To determine the stationarity of the EOG activity, Hilbert envelopes of each subject's five activity signals (normal glance, left eye closed, right eye closed, frowning and eyebrows up) are extracted and divided into non-overlapping windows of one second each. This allows to obtain the mean and standard deviation of the signal windows. The derived feature sequences are linearized by taking the logarithm and average (AVG) over all subject sequences for each activity, independently of trending (sequences are visualized in \autoref{fig:stdavg_envelopes_among_ocularactivities}. These averaged sequences are then subjected to the Augmented Dickey-Fuller (ADF) test \citep{b24.1}. As can be seen in \autoref{tab:ADF test}, the null hypotheses of having unit roots are rejected in normal glance, frowning and eyebrows up activities (p < .5) for both attributes. However, unit roots are detected in both STD and AVG feature sequences for left eye closed and right eye closed activities (left eye closed: p = .651 for STD, p = .879 for AVG; right eye closed: p = .464 for STD, p = .844 for AVG). 


\begin{table*}[width=.9\textwidth]
\caption{Augmented Dickey-Fuller test for signal stationarity.}\label{tab:ADF test}
\begin{tabular*}{\tblwidth}{c@{\hskip 1.2in}c@{\hskip 1.2in}c LLLLLLL@{} }
\toprule
\multirow{2}{*}{Ocular activity}                                 & \multicolumn{2}{c}{ADF stationarity test results} \\ \cline{2-3}
& STD of Hilbert envelopes    & AVG of Hilbert envelopes \\
\midrule
Normal glance            & p < .05             & p < .05        \\
Left eye closed          & p = .651            & p = .879         \\
Right eye closed         & p = .464            & p = .844         \\
Frowning                 & p < .05             & p < .05        \\
Eyebrows up              & p < .05             & p < .05        \\
\bottomrule
\end{tabular*}
\end{table*}


\subsection{Statistical comparisons among Hilbert envelopes}
\vspace{1mm}

In this step, both the STD and AVG attribute sequences are simplified by averaging every five consecutive values in each long attribute sequence as a time-independent series. It is taken into account that they are static activities. Next, the sequences are subjected to repeated measures ANOVA (RM-ANOVA) with a within subject design. Test results are shown in \autoref{tab:RM_AOVA_testresults}. The sphericity (describing the situation in which all combinations of the related groups have equal variances in their differences) of the sequences is conditioned by Greenhouse-Geisser correction (STD of Hilbert envelope sequences: p = .414; AVG of Hilbert envelope sequences: p = .270). The RM-ANOVA test results of the corrected sequences showed no significance for either attribute (STD of Hilbert envelope sequences: p = .114; AVG of Hilbert envelope sequences: p = .421). thus, there is no possibility to distinguish activities that are not correlated in the distribution of envelope variations by time-domain attributes alone, and further handcrafted feature extraction in both time and frequency-domains is vital.

\begin{table*}[width=.9\textwidth]
\caption{Repeated measures ANOVA comparison within-subject.}\label{tab:RM_AOVA_testresults}
\begin{tabular*}{\tblwidth}{c@{\hskip 0.6in}c@{\hskip 0.7in}c LLL@{} }
\toprule
Feature                                  & Test of sphericity    & Greenhouse-geisser corrected p-value \\
\midrule
STD of Hilbert envelope sequences         & p = .414             & p = .114        \\
AVG of Hilbert envelope sequences          & p = .270             & p = .421         \\
 
\bottomrule
\end{tabular*}
\end{table*}

\subsection{Feature engineering}
\vspace{1mm}

In order to increase the classification performance, the attributes of each of the three windows next to each other are combined to provide a time-domain-dependent semantic context. Attributes are added end to end as one-dimensional and a new attribute with a total of 87 dimensions has emerged. However, in order to be able to classify with the convolutional neural network, a two-dimensional feature input is created by adding the adjacent feature sequences one under the other.

\subsection{Classifier Selection}
\vspace{1mm}

A total of 16 different classifiers are chosen: 2D Convolutional Neural Networks (CNN) \citep{b22.1}, Long-short term memory (LSTM) \citep{b22.2}, 1D CNN with LSTM, 2D CNN with LSTM, Linear discriminant analysis (LDA) with shrinkage value, K-nearest neighbours (KNN), Gaussian process classifier (GPC) \citep{b22.3} based on Laplace approximation, Bagging classifier \citep{b22.4}, Adaboost  \citep{b22.5}, support vector classifier (SVC), XGBoost  \citep{b22.6}, Random forest, Lightgbm (LGBM)  \citep{b22.7}, Voting ensemble classifier, and a stack ensemble meta-model \citep{b22.8}. The stack ensemble uses stacking as a tool for meta-model training by combining the results from independently trained classifiers (KNN, GPC, Bagging, Adaboost, SVG, SGD, XGBoost, Random Forest, and LGBM). It uses the stochasticity of the models' individual predictions, and the meta-model makes a final decision from incoming inputs. A voting classifier with a soft voting mechanism is used to predict the argmax of total probabilities. Sklearn \citep{b23} and Keras \citep{b23.1} packages are used to execute the algorithms in a Python3.9 environment. 

\subsubsection{Recursive feature elimination}
\vspace{3mm}
The redundant features are removed from the entire space using recursive feature elimination with 10-fold cross validation (RFECV) \citep{b23.1.1}. Due to its suitability for deriving feature significance coefficients for each feature, the decision tree classifier \citep{b23.12} is employed. 

\subsubsection{Hyperparameter optimization and selection}
\vspace{3mm}

For all classifiers, the optimum hyperparameters are discovered in the initial step using hyperparameter optimization. Alternative optimization techniques are designed to identify a relatively decent hyperparameter combination rather than a grid search. The later attempts all hyperparameter combinations to find the best one, resulting in a very high computation cost. Instead, a Bayesian hyperparameter optimization method is used to optimize the model parameters, except the deep learning models. This method is chosen because it is computationally much cheaper and converges quickly to high classification performance \citep{b23.2}. In all four different deep learning models, Adamax \citep{b23.3} is chosen as the optimizer. Dropout (50\%) is added between each layer to ensure regularization. The rectifier linear unit (reLu) is chosen as the activation function of each layer, and BatchNormalization is used to standardize sample chunks that are divided into mini-batches after each reLU activation. 

\subsubsection{Leave-one-session-out cross validation results}
\vspace{3mm}

The whole dataset is divided into six folds according to leave-one-session-out cross validation (LOSO) \citep{b23.31}. Corresponding to this, any five of the six subjects are trained, and the remaining subject is used as a test. This prevents data leakage from the test subject into the training dataset, thus produces performance results that are substantially more realistic than those of the traditional k-fold cross validation with shuffling. Within each fold, the training data are first normalized and the test data (test subject) are normalized according to the mean and standard deviation data calculated for each feature. The next step is to apply RFECV to the training data to determine the importance of the features in each subset and simplify the test dataset according to the identified features. After this process, again only the training set is subjected to hyperparameter optimization and the test data are isolated from this process as well.

The LOSO performances of each model with and without harmonic filtering of EOG signals are shown in \autoref{tab:classificationResults}. The evaluation metrics employed include mean accuracy (with standard deviation based on individual fold performances), precision, recall, and f1-score. The average of all test results are recognized as the final performance results of the models. Since the sample sizes of the classes are equal, the accuracy metric is more decisive than the others. Accordingly, 2D CNN provided the highest accuracy (72.35\%). Random forest and stacked ensemble models are close behind with 72.24\% and 72.19\% respectively. The other deep learning models (LSTM, 1D CNN+LSTM and 2D CNN+LSTM) give relatively lower performances (71.43\%, 67.80\% and 66.88\% respectively). Voting ensemble, LGBM, XGBoost, SVG, Bagging, KNN, SGD, GPC and Adaboost models give accuracy of 69.81\%, 69.32\%, 69.32\%, 68.72\%, 67.26\%, 58.77\%, 58.66\%, 57.79\% and 33.17\% respectively. The performance of all models for EOG signals without harmonic filtering are noticeably lower. Random forest model showed 23.17\% lower LOSO accuracy performance than the one evaluated with the harmonic filtered EOG data. Similarly, when testing  the other models without harmonic filtering, the stacked ensemble and 2D CNN models perform 9.17\% and 11.68\% worse, respectively. The results are presented extendedly in \autoref{tab:classificationResults}. 

\begin{figure}[htbp]
 \centering
 \includegraphics[width=1\columnwidth,keepaspectratio=true]{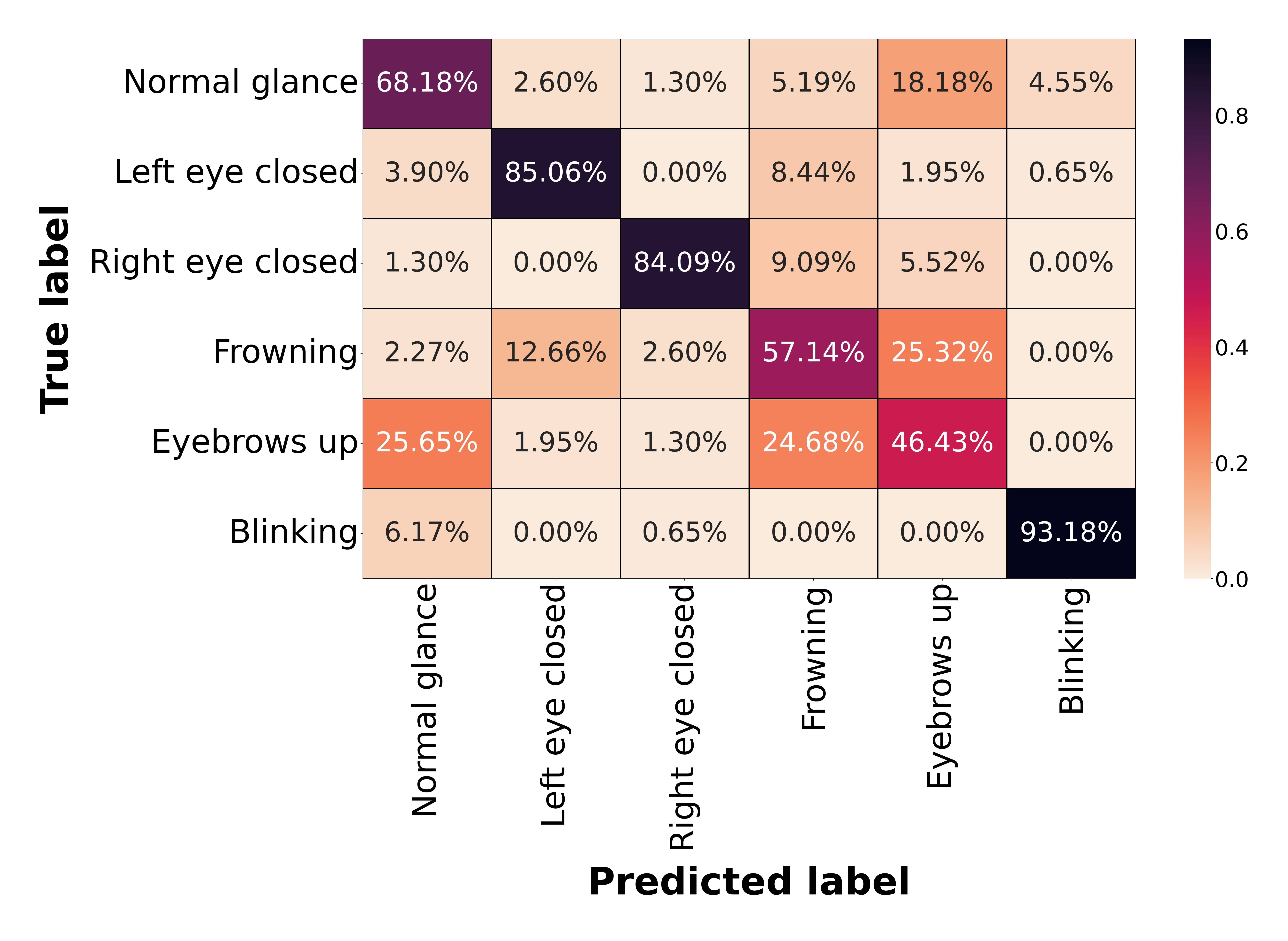}
 \caption{Normalized Confusion Matrix of overall 2D CNN LOSO performance}
  \label{fig:confusionMatrix}
\end{figure}

In addition, the normalized confusion matrix of the model with average 2D CNN (the highest performing model) LOSO scores is extracted and shown in \autoref{fig:confusionMatrix}. Accordingly, eye blink activity is by far the best predicted class (93.18\%). This is followed by right and left eye closed activities (85.06\% and 84.09\%). Normal glance and frowning yielded considerably lower classification results (68.18\% and 57.14\%). Eyebrows up achieved the lowest classification success with a performance of 46.75\%. 

\begin{table*}[width=\textwidth]
\caption{LOSO classification results.}\label{tab:classificationResults}
\begin{tabular*}{\tblwidth}{c@{\hskip 0.07in}c@{\hskip 0.0in}cccccccc LLLLLLLLLLLLLLLLLL@{} }
\toprule
\multirow{2}{*}{Model}  & \multicolumn{4}{c}{With harmonic filtering} & & \multicolumn{4}{c}{Without harmonic filtering}\\ \cline{2-5}  \cline{7-10}
                        & Accuracy    & Precision & Recall & F1 score &          & Accuracy & Precision & Recall & F1 score\\
\midrule
\emph{\textbf{2D CNN }}              & \textbf{0.7235 $\pm$ \ 0.0898}  & \textbf{0.7485} & \textbf{0.7224} &\textbf{0.7042} & & \textbf{0.6331 $\pm$ \ 0.0930}   & \textbf{0.6317} & \textbf{0.6103} & \textbf{0.5757}\\
\emph{LSTM}                 & 0.7143 $\pm$ \ 0.1032   & 0.7246 & 0.7143 & 0.6976 & & 0.6103 $\pm$ \ 0.0830    & 0.6340 & 0.6119 & 0.5735\\
\emph{1D CNN + LSTM}        & 0.6780 $\pm$ \ 0.1055    & 0.6966 & 0.6780  & 0.6627 & & 0.5872 $\pm$ \ 0.0537   & 0.6040 & 0.5872 & 0.5600\\
\emph{2D CNN + LSTM}        & 0.6688 $\pm$ \ 0.1059   & 0.6985 & 0.6688 & 0.6582 & & 0.5561 $\pm$ \ 0.0893    & 0.5764 & 0.5561 & 0.5338\\
\emph{LDA with shrinkage}   & 0.5774 $\pm$ \ 0.1939   & 0.6503 & 0.5774 & 0.5777 & & 0.5717 $\pm$ \ 0.1557    & 0.6262 & 0.5717 & 0.5573\\
\emph{KNN}                  & 0.5877 $\pm$ \ 0.0883   & 0.6183 & 0.5877 & 0.5758 & & 0.4829 $\pm$ \ 0.0979    & 0.5578 & 0.4829 & 0.4721\\
\emph{GPC}                  & 0.5779 $\pm$ \ 0.0260    & 0.6228 & 0.5779 & 0.5641 & & 0.4764 $\pm$ \ 0.0809    & 0.6496 & 0.6331 & 0.6113\\
\emph{Bagging}              & 0.6726 $\pm$ \ 0.0517   & 0.6626 & 0.6726 & 0.6517 & & 0.5783 $\pm$ \ 0.0864    & 0.6418 & 0.6103 & 0.5865\\
\emph{Adaboost}             & 0.3317 $\pm$ \ 0.0234   & 0.3021 & 0.3317 & 0.2520 & & 0.4653 $\pm$ \ 0.1757    & 0.6262 & 0.5717 & 0.5573\\
\emph{SVG}                  & 0.6872 $\pm$ \ 0.0807   & 0.7212 & 0.6872 & 0.6704 & & 0.5377 $\pm$ \ 0.0750     & 0.5578 & 0.4829 & 0.4721\\
\emph{SGD}                  & 0.5866 $\pm$ \ 0.1406   & 0.6232 & 0.5866 & 0.5743 & & 0.5508 $\pm$ \ 0.0761    & 0.5544 & 0.4764 & 0.4668\\
\emph{XGBoost}              & 0.6932 $\pm$ \ 0.0417   & 0.6787 & 0.6932 & 0.6652 & & 0.6000 $\pm$ \ 0.0812    & 0.5846 & 0.5783 & 0.5481\\
\emph{\textbf{Random forest}}           & \textbf{0.7224 $\pm$ \ 0.0711}   & \textbf{0.7176} & \textbf{0.7235} & \textbf{0.7010} & & \textbf{0.5907 $\pm$ \ 0.1135}    & \textbf{0.4907 } & \textbf{0.4653} & \textbf{0.4404}\\
\emph{LGBM}                 & 0.6932 $\pm$ \ 0.0613   & 0.6919 & 0.6932 & 0.6723 & & 0.6066 $\pm$ \ 0.0742    & 0.5837 & 0.5377 & 0.5098\\
\emph{Voting ensemble}           & 0.6981 $\pm$ \ 0.0723   & 0.6937 & 0.6981 & 0.6753 & & 0.6103 $\pm$ \ 0.0993    & 0.5743 & 0.5508 & 0.5234\\
\emph{\textbf{Stacked ensemble}}     & \textbf{0.7219 $\pm$ \ 0.0655}  & \textbf{0.752}  & \textbf{0.7219} & \textbf{0.6987} & & \textbf{0.6119 $\pm$ \ 0.0987}       & \textbf{0.6062} & \textbf{0.6000} & \textbf{0.5665}\\
\bottomrule
\end{tabular*}
\end{table*}


\subsection{Clustering for the EOG data grouping exploration}
\vspace{3mm}
As seen in the confusion matrix, normal glance, frowning and eyebrows up activities are considerably confused with each other. This poses a new problem, to what extent the ocular activities can be differentiated from each other based on the extracted features.

\subsubsection{Assessment of dimension reduction methods}
\vspace{3mm}
In order to cluster the feature set more effectively, six different dimension reduction algorithms (Principal component analysis (PCA), Incremental PCA  (IPCA)\citep{b23.4}, Fast independent component analysis  (fICA)\citep{b23.5}, Sparse PCA (sPCA) \citep{b23.6}, Truncated singular value decomposition (tSVD)  \citep{b23.7} and t-distributed Stochastic Neighbor Embedding (t-SNE) \citep{b23.8}) are used to transform the 87 dimensional feature set to 2 dimensions. The resulting datasets are clustered using K-means \citep{b23.9} and Fuzzy C-means \citep{b23.10} algorithms. Four different evaluation metrics: homogeneity, completeness, V-measure and silhouette scores, are used to evaluate the clustering performance of the dimension reduction algorithms \citep{b23.11}. As can be seen in \autoref{tab:2D_clustering_results}, t-SNE shows the highest performance in all metrics by a clear margin. Based on the silhouette metric, the closer the score is to 1, the more successful it is to cluster the data set in the recommended number of groups. Based on the t-SNE algorithm, AWS score of K-means gives 0.5901 and Fuzzy C-means gives 0.5573. The silhouette scores of the other metrics (PCA, IPCA, fICA, sPCA, tSVD and MBDL) for K-means are: 0.4731, 0.4779, 0.4678, 0.4697, 0.4727 and 0.4668; and for Fuzzy C-means: 0.4389, 0.4024, 0.4366, 0.4031, 0.4102 and 0.4386. 

\begin{figure}
     \centering
     \resizebox{\columnwidth}{!}{%
     \begin{subfigure}[t]{\columnwidth}
         \centering
         \includegraphics[width=\columnwidth]{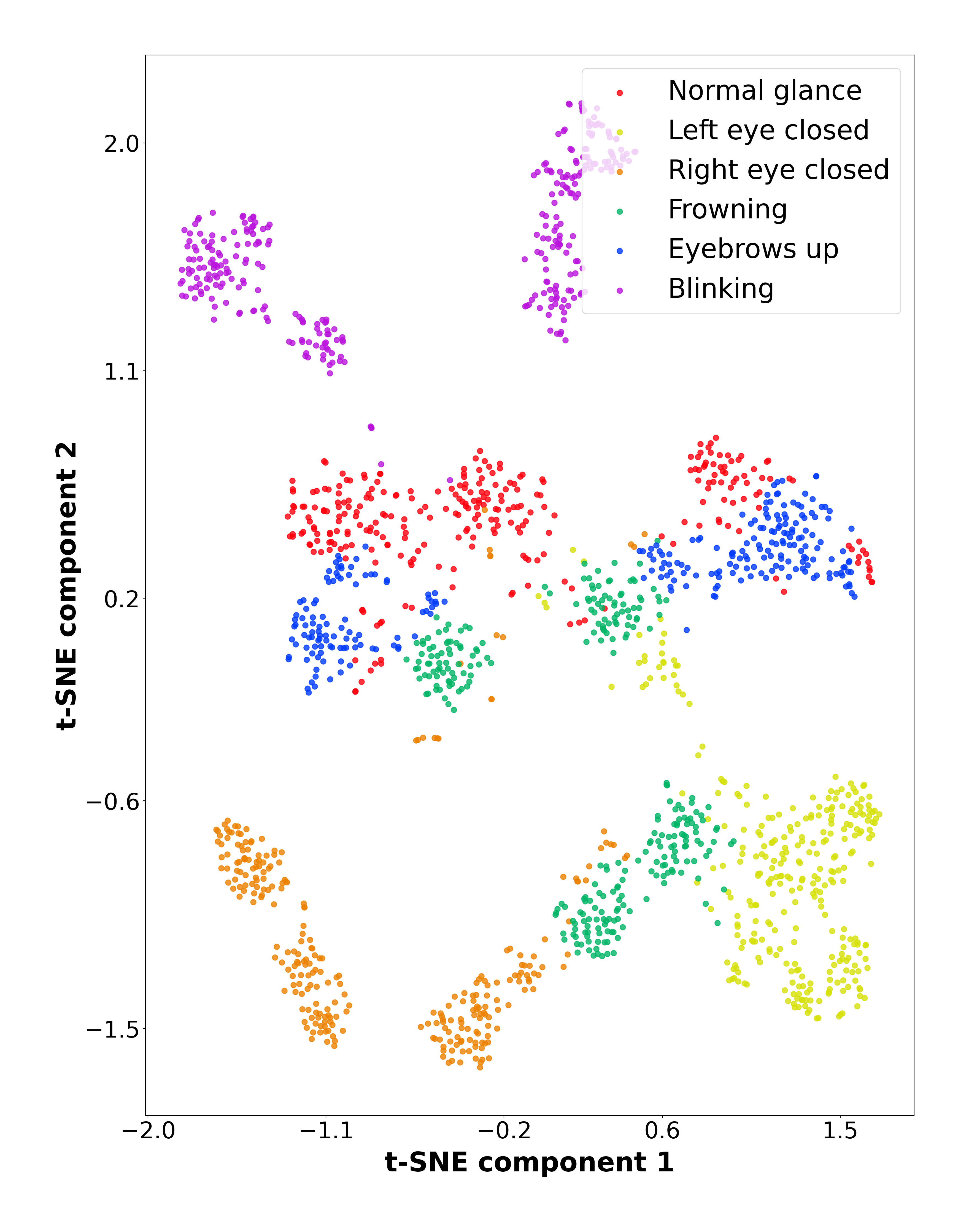}
     \end{subfigure}%
     }
     \caption{Visualization of class distribution with t-SNE. It can be observed that the little sub-groups of frowning, normal glance, and eye brows up occupy a position that can mingle with each other in the centre of the figure, despite the fact that the classes of frowning, eyebrows up, and normal look overlap each other. Through this vantage point, it is clear why the classes for the normal  glance, frowning, and eyebrows up activities in the confusion matrix are mixed together.}
     \label{fig:visualization_of_class_distribution_with_tSNE}
\end{figure}

In a nutshell, all the dimension reduction algorithms aside from t-SNE fail to turn the dataset into a grouping state, at least partially. To examine the distribution of ocular activity, the t-SNE components are examined as seen in \autoref{fig:visualization_of_class_distribution_with_tSNE}. Upon a closer inspection, it seems that the values for the normal gaze, frowning, and raised eyebrow classes are partially mingled together in the middle of the figure.   

\subsubsection{Optimum number of cluster evaluation}
\begin{figure}
     \resizebox{\columnwidth}{!}{%
     \begin{subfigure}[t]{\columnwidth}
         \centering
         \includegraphics[width=\columnwidth]{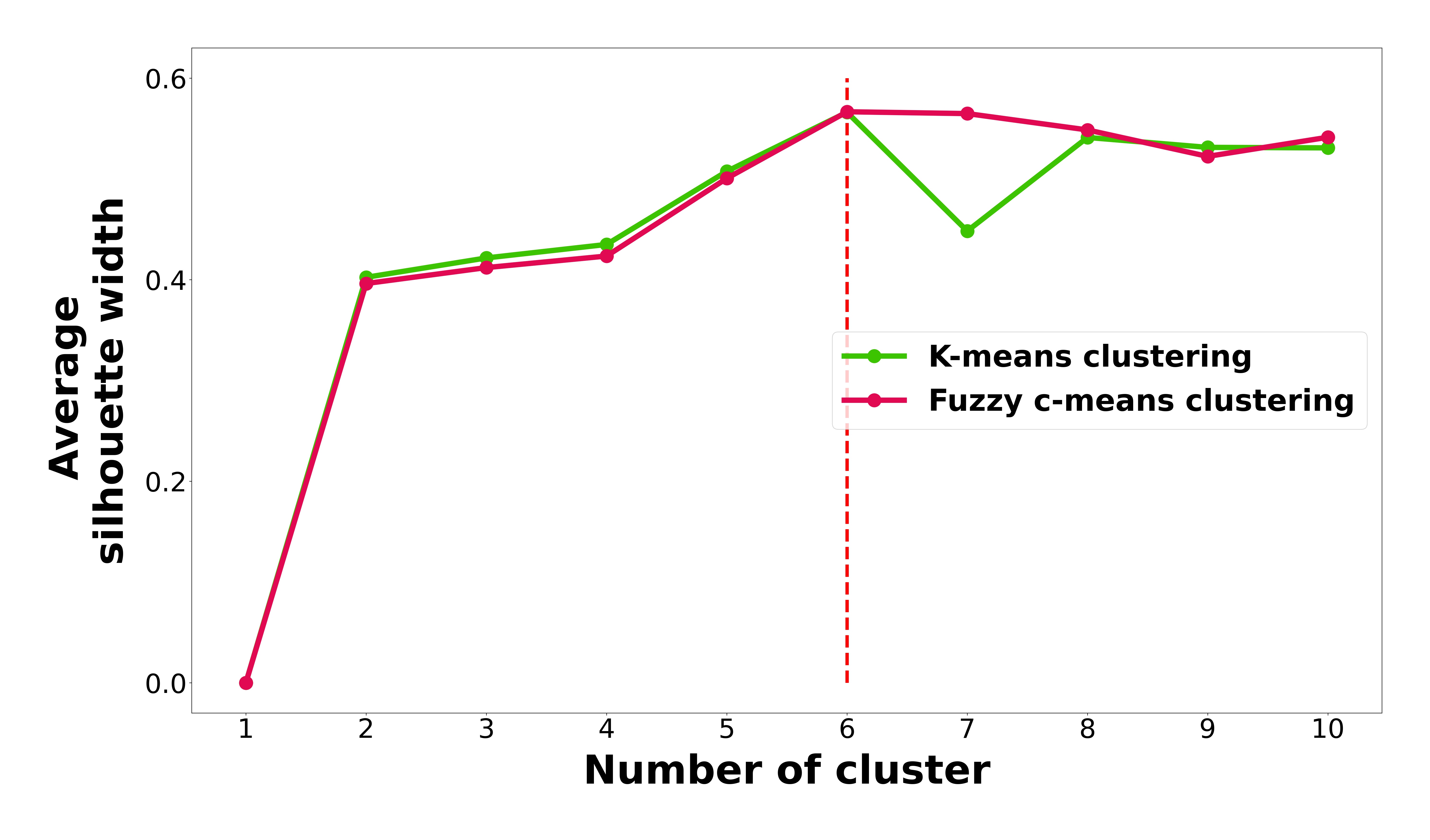}
     \end{subfigure}
     }
     \caption{Evaluation of optimum number of clusters for both K-means and Fuzzy C-means clustering algorithms. Average silhouette width is the metric used for assessing clustering performance for the number of clusters between 2 to 10. The optimum number of clusters for both algorithms is found to be 6.}
     \label{fig:optimum_number_of_clusters}
\end{figure}

\vspace{3mm}
The ideal number of clusters is further examined empirically. The t-SNE component set is once more clustered using the K-means and Fuzzy C-means clustering algorithms according to all cluster indices between 2 and 10. The ideal cluster index is discovered by assessing each of them in accordance with the average silhouette width metric (ASW) \citep{b23.1.2}. The average distance between data points within a cluster is compared to the average distance between data points within other clusters to calculate the ASW, which is an intuitive and basic evaluation of cluster quality. The formula of ASW is shown in Eq.~(\ref{eq:ASW}). The maximum ASW values are found to be 0.5662 and 0.5669 for K-means and Fuzzy C-means respectively when there are 6 clusters(as shown in \autoref{fig:optimum_number_of_clusters}). 

\begin{equation}
\begin{split}
\label{eq:ASW}
S_{C, x, d(i)} = \frac{b_{d(i)} - a_{d(i)}}{max(a_{d(i)} - b_{d(i)})} \\ 
S_{C, x} = \frac{1}{n}\sum_{k=1}^{n}S_{C,x,d_{n}} \\ 
S_{C} = \frac{1}{x}\sum_{k=1}^{x}S_{C,x}
\end{split}
\end{equation}

\noindent
$d_i$ are the data points for a given cluster $x$, $a_{d(i)}$ is an average distance between a point $d_{(i)}$ and all other data points in the same cluster, $b_{d(i)}$ is the average distance of $d_{(i)}$ to the closest cluster excluding the cluster to which the point belongs to. $S_{C, x}$ is an average silhouette score of a given cluster, and $S_{C}$ is a grand average silhouette width of mean silhouette scores of all clusters for a clustering algorithm grouped with the size of $C$. 


\begin{table*}[width=\textwidth]
\caption{Clustering performance evaluation of various dimension reduction algorithms}\label{tab:2D_clustering_results}
\begin{tabular*}{\tblwidth}{cccccccc LLLLLLLLLLLLL@{} }
\toprule
\multirow{2}{*}{}      & \multicolumn{4}{c}{K-means clustering results} &  & \multicolumn{4}{c}{Fuzzy C-means clustering results} \\ \cline{2-5}  \cline{7-10}
Model & Completeness & Homogeneity & Silhouette & V-measure & & Completeness & Homogeneity & Silhouette & V-measure\\
\midrule
\emph{PCA}            &  0.4703       &    0.3874   &    0.4731  &    0.4248  & & 0.4218 & 0.3645 & 0.4389 &    0.3911\\
\emph{IPCA}           &  0.4705       &    0.3849   &    0.4779  &    0.4234  & & 0.4323 & 0.4013 & 0.4024 & 0.4162\\
\emph{fICA}           &  0.4512       &    0.3680   &    0.4678  &    0.4054  & & 0.4061 & 0.3471 & 0.4366 & 0.3743\\
\emph{sPCA}           &  0.4528       &    0.3696   &    0.4697  &    0.4070  & & 0.4285 & 0.3965 & 0.4031 & 0.4119\\
\emph{tSVD}           &  0.4694       &    0.3871   &    0.4727  &    0.4242  & & 0.4368 & 0.4082 & 0.4102 & 0.4220\\
\emph{\textbf{t-SNE}}  & \textbf{0.6133} & \textbf{0.5901} & \textbf{0.5583}  & \textbf{0.6015} & & \textbf{0.6059} & \textbf{0.5836} & \textbf{0.5573}  & \textbf{0.5945}\\
\bottomrule
\end{tabular*}
\end{table*}

\subsection{Real-time Experiments}
\vspace{3mm}

\begin{figure}[htbp]
 \centering
 \includegraphics[width=1\columnwidth,keepaspectratio=true]{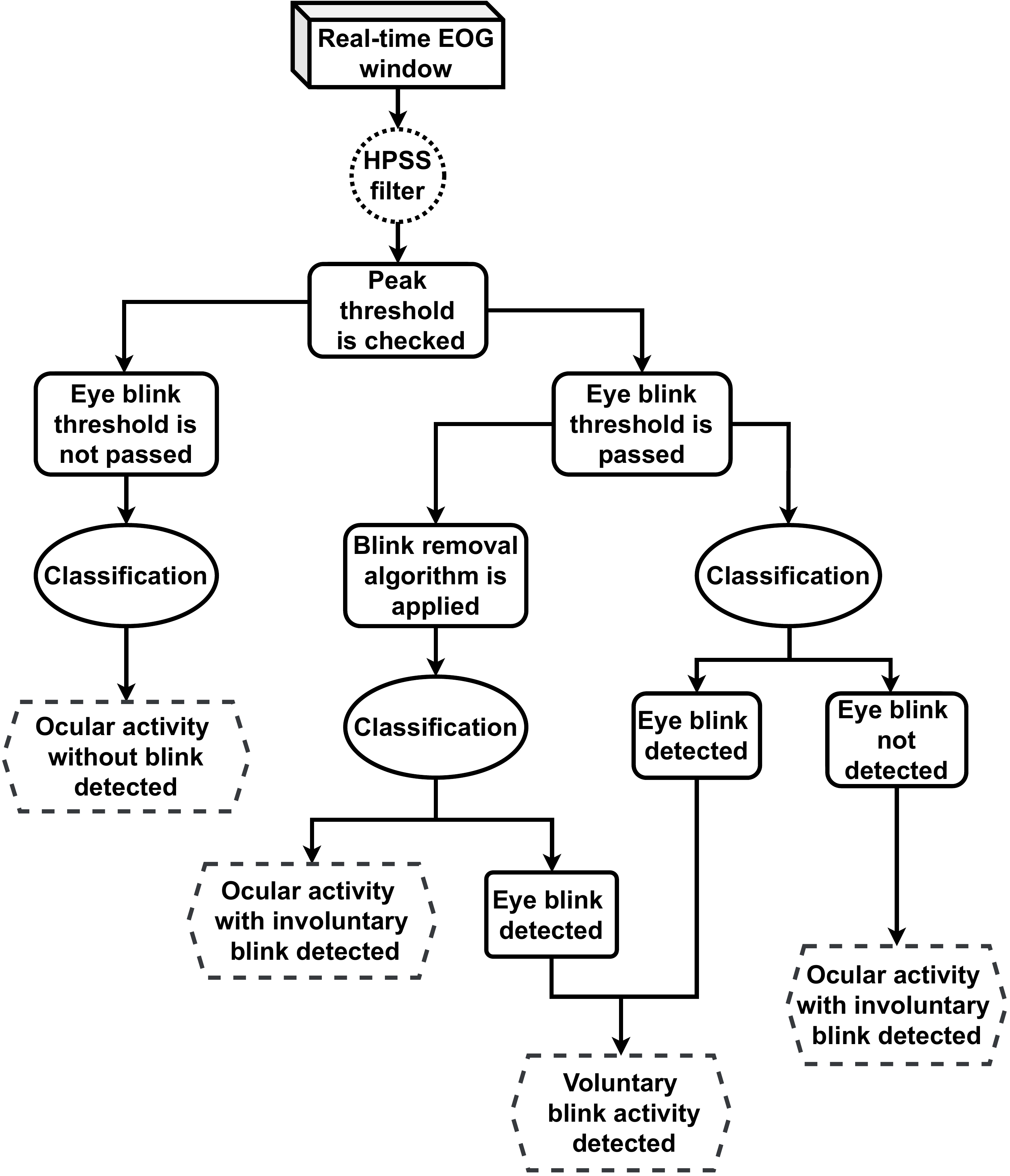}
 \caption{Real-time activity prediction.}
  \label{fig:activityAssessment}
\end{figure}

During real-time testing, a rule-based evaluation system is created to identify whether blinking movements are spontaneous or reflex and the associated logic diagram is shown in \autoref{fig:activityAssessment}. In order to check the peak threshold, an HPSS filter is applied to the real-time EOG window. Signals that surpass the threshold are then classified independently, both with and without the use of the blink removal technique. It is considered a voluntary eye blink if the classifier detects blink activity in both predictions. However, if the signal predicts an activity other than eye blink after the remove artifact is removed, it is predicted as one of the five activities (normal glance, left eye closed, right eye closed, frowning, eyebrows up) that contain involuntary eye blink. If the peak threshold is not detected, the blink removal algorithm is not applied and the classifier predicts the activity.

\begin{figure}[htbp]
 \centering
 \includegraphics[width=1\columnwidth,keepaspectratio=true]{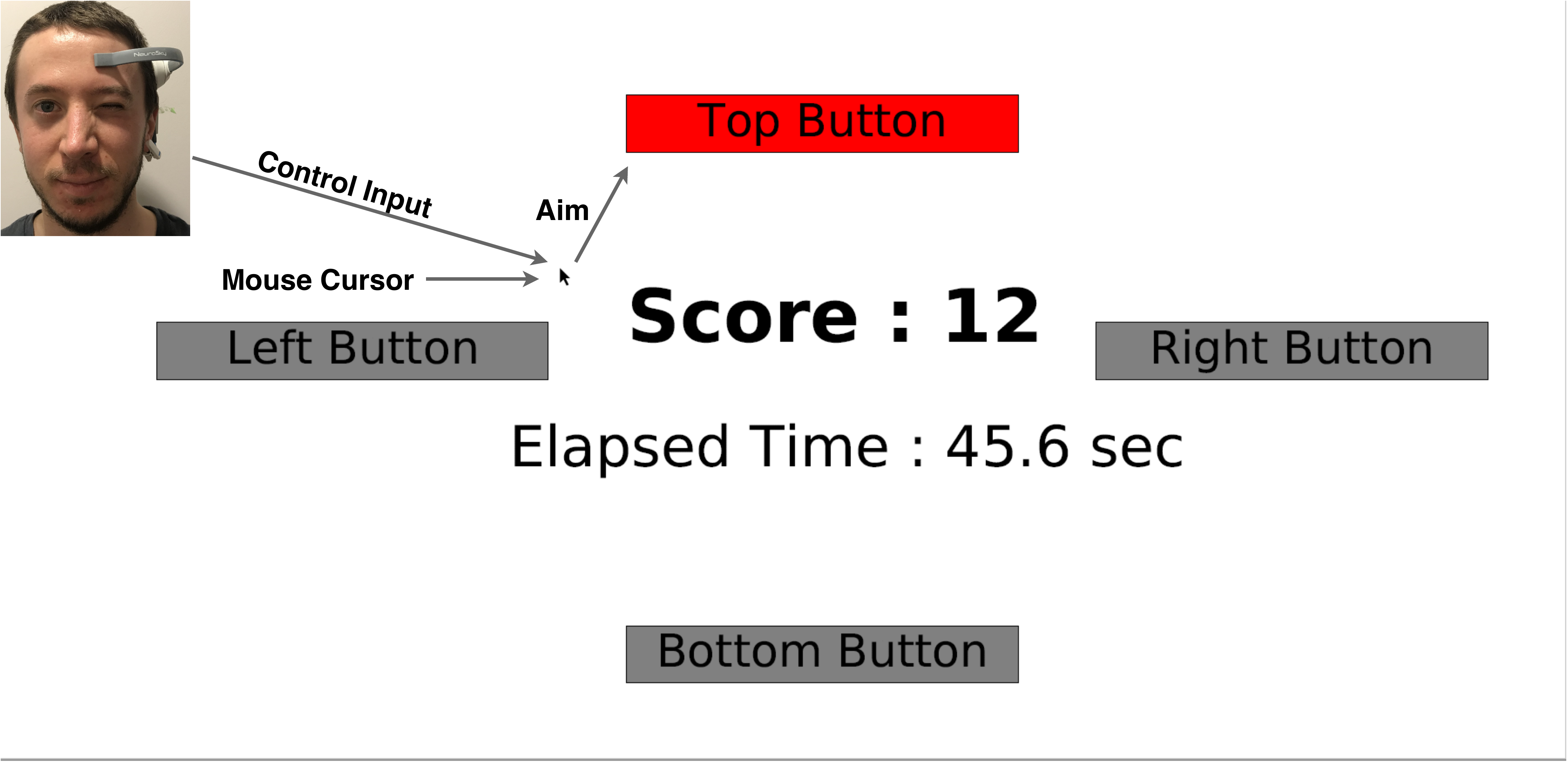}
 \caption{Real-time test GUI}
  \label{fig:realtimeTestGUI}
\end{figure}

The GUI is implemented to control the mouse cursor in real time. The GUI has four buttons, one on the left, one on the right, one on top, and one at the bottom. The system automatically assigns a button for the subject to navigate the mouse cursor to and hit this red colored button. Gray-colored buttons are not meant to be clicked during the task duration. In the middle of the GUI, there is a score. Each time a participant clicks the red-colored button, a score rises by one point. The GUI is shown in \autoref{fig:realtimeTestGUI}. 

A GUI built in the Python environment is also used to test the system in real time. Subjects are asked to click on one of four buttons positioned at the margins of the GUI, which the system selects for them automatically. Random forest classifier is used for real-time prediction due to low computational complexity compared with the 2D CNN model, and having the second best LOSO performance. All subjects are given a five-minute test with the GUI, and their scores as well as the time it took them to obtain those scores are recorded. Finally, all participants are invited to fill in a survey and assess the system's real-time efficiency and convenience.  Scoring is limited between 0-4 and average scores of the survey are shown in \autoref{tab:survey}.

\begin{table}[width=.9\linewidth,cols=4,pos=h]
\caption{Survey of real-time system assessment.}\label{tab:survey}
\begin{tabular*}{\tblwidth}{@{} LLLL@{} }
\toprule
Subjects & Score & Efficiency & Convenience\\
\midrule
Subject 1 &  45   & 3.2  & 2.9 \\
Subject 2 &  21   & 2.4  & 2.1 \\
Subject 3 &  66   & 3.3  & 3.2 \\
Subject 4 &  54   & 3.5  & 3.8 \\
Subject 5 &  33   & 2.8  & 2.7 \\
Subject 6 &  58   & 3.7  & 3.5 \\
Average   & 46.16 & 3.15 & 3.03 \\
STD       & 16.75 & 0.47 & 0.605 \\
\bottomrule
\end{tabular*}
\end{table}

\begin{figure*}
     \centering
     \resizebox{\columnwidth}{!}{%
     \begin{subfigure}[t]{0.4\textwidth}
         \centering
         \includegraphics[width=0.8\columnwidth]{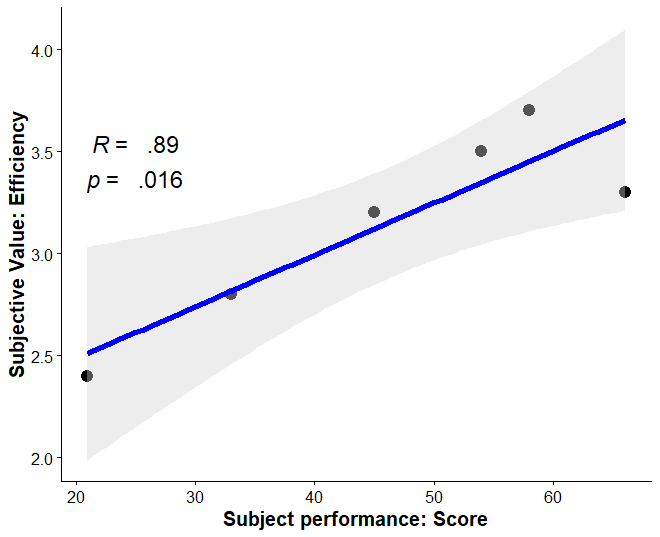}
         \caption{}
         \label{fig:efficiencyScore}
     \end{subfigure}%
     }
     \hfill
     \resizebox{\columnwidth}{!}{%
     \begin{subfigure}[t]{0.4\textwidth}
         \centering
         \includegraphics[width=0.8\columnwidth]{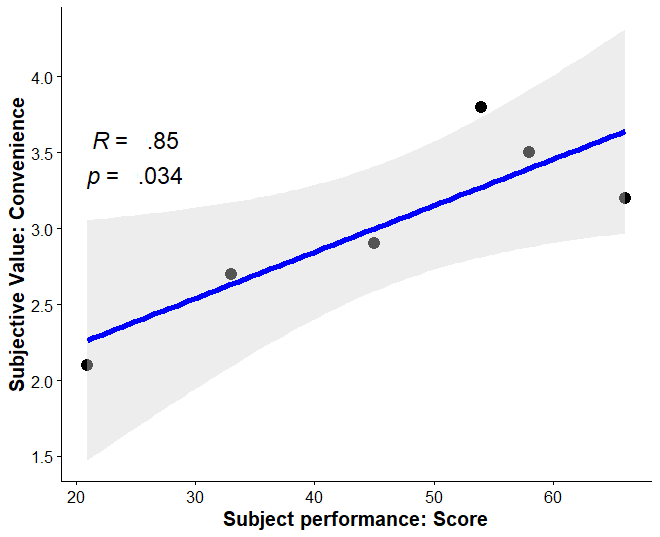}
         \caption{}
         \label{fig:convenienceScore}
     \end{subfigure}%
     }
     \caption{Pearson correlation between subjective values and scores. Pearson correlation test between the subject performance pointed out by their 'score' and their subjective evaluation of the interface indicated by 'efficiency' in \textbf{(a)}. The test yielded a significant positive correlation R = 0.89 and p = .016. This result confirms the validity of the efficiency values as they are in direct correlation with the performance results of the subjects. This confirmation suggests that the participants found the interface efficient. Moreover, Pearson correlation test between the subject performance pointed out by their 'score' and their subjective evaluation of the interface indicated by 'convenience' \textbf{(b)}. The correlation  The test yielded a significant positive correlation R = 0.85 and p = .034. Since the convenience values and the respondents' performance outcomes are directly correlated, this result supports the validity of the convenience values. This result suggests that the participants found the interface convenient, thus they would choose to use it long-term.}
     \label{fig:pearsonCorrelation}
\end{figure*}

According to the real-time tests, the average score and standard deviation (STD) of the six subjects are found to be 46.16 and 16.75 respectively. The standard deviation value demonstrates a considerable performance variation between the participants, which is considerable when compared to the mean value. Accordingly, average efficiency and convenience are 3.15 and 3.03; STD of values are 0.47 and 0.605, respectively. In addition, a Pearson correlation test is performed to determine the relationship between the subjects' scores and the survey results. Both efficiency and convenience subjective values have a statistically significant positive correlation (p=0.016 for R=0.89 and p=0.034 for R=0.85) with the test scores (\autoref{fig:pearsonCorrelation}). The positive correlation between the GUI performance scores and questionnaire values suggests that users are likely to use this device in their daily tasks. However, it is still observed subjects with below-average scores have trouble efficiently using the up-down commands. 

\section{Conclusion}

The goal of this research is to explore the dynamics of the ocular activities using minimal sensing EOG device and multiple strategies in a data science manner; and to provide a reliable and simple HMI-based system that will aid in controlling the mouse cursor activity in an alternative way. The pipeline is primarily directed at patients with partial or close to complete motor disabilities, including LiS patients, apart from people suffering from ALS and MS. However, all these patients have one common characteristic that manifests itself at a very high rate - even though they are (almost) paralyzed, they do not lose their ability to control their eyes. Hence, EOG-based HMI solutions can be of great aid. Closing the left or right eye, frowning, lifting eyebrows, eye blinking, and neutral glancing are among the six ocular activities estimated by the developed system. Corresponding computer commands aim to navigate the mouse cursor for full scale 2D control, including single and double clicking (eye blinking twice in a row).

Given that EOG activities are artifactual markers in comparison to EEG, a proper data cleaning protocol is inapplicable because the main components of ocular activities are themselves artifacts. Excessive preprocessing, in this case, removes those meaningful artifacts. However, these are essential for an EOG-based HMI pipeline to extract the corresponding features for contrasting the eye activities. Consequently, apart from high-pass filtering, notch filtering, and blink removal algorithm, no further data cleaning is applied. The majority of EOG-HMI based studies use a band-pass filter between 0.1 - 30 Hz \citep{b12.1} and are unable to utilize higher frequency harmonic information. Nonetheless, by removing the percussive components from the basic EOG signal, harmonic signals in higher frequency bands are recovered. This results in a novel representation that more precisely reflects ocular activities. Harmonic signals are filtered out of a signal's percussive components and applied in the frequency domain to keep only the energies of the harmonic spectrum. Using harmonic filters in the spectral domain of EOG signals and cleaning non-periodic artifacts from raw signals provides remarkably more information about eye actions in the recorded data. This is found to be more effective than using only high-pass filtered raw data by comparing LOSO validation results.

Initially, the Hilbert envelopes of the signals are windowed and the simplified EOG sequences of each activity are extracted by calculating the average and standard deviation features (as energy variation indicators). Repeated measures ANOVA is used for within-subject comparison of each activity and no statistical difference is found between ocular activity sequences. Additionally, the stationarity of the signals is checked to see if there is any unit root in the linear plane for each activity. The results of the ADF test indicate that the left eye closed and right eye closed sequences have a certain unit root (non-stationary) even though they are static activities. In the face of these unexpected outcomes, it is speculated that the eye activities, which are thought to be static, perhaps are caused by micro-blinks in the time-direction due to cumulative fatigue. However, if an interactive system is to be controlled with an HMI-EOG, it is not possible to avoid static activities (e.g. moving the mouse cursor to the left for a certain period of time, other repetitive process-oriented activities), but it is recommended for time-dependent dynamic twitching of the ocular muscles to be taken into account in future studies.

In the next step, various feature extraction algorithms are performed to create hand-crafted features from harmonically filtered EOG signals. In order to increase the classification performance, 29 features are extracted from each windowed signal and the feature space of the samples is tripled (87 features) by merging the features of each three adjacent windows with a padding of 500 ms. Sixteen different classification algorithms are trained separately both with and without the harmonic filtered feature sets. All the algorithms classified the harmonic filtered features significantly better due to the increased efficiency of the harmonic components in the EOG data as opposed to their diffuse percussive energy. The performance of the algorithms is evaluated according to the LOSO protocol. The 2D CNN model achieved the highest accuracy performance, predicting ocular activities at a rate of 72.35\%. Additionally, the average confusion matrix is extracted from the LOSO evaluation of 2D CNN model which indicates that the best performing classifier is having trouble in distinguishing well between frowning and eyebrows up, and (although to a lesser extent) the normal glance. This happens even though these facial expressions are being manifested by different facial muscles \citep{b30}.

In the last step, the distributions of the total feature set and the overlap of the features according to the classes are analyzed as a result of the normal glance, frowning, and eyebrows up activities intermingled in the confusion matrix. The 87-dimensional feature collection is reduced to 2 dimensions using six distinct unsupervised dimension reduction approaches. Visual inspection reveals that the t-SNE algorithm transforms the feature set in a way that distributes the feature set classes in the most discriminatory way. However, it is clear that the typical acts of glancing, frowning, and raising the eyebrows partially overlap. In addition to this, an empirical analysis of a generalizable grouping is conducted. Different cluster numbers are tested with K-means and Fuzzy C-means algorithms, and AWS results showed that the total feature set is relatively optimally divided into six clusters for both algorithms. The maximum AWS values are around 0.56 for both clustering techniques, indicating that presently superior clustering performance cannot be achieved.

In the future study, real-time spectrogram images can be trained with deep learning algorithms to predict more comprehensive ocular activities, in addition to handcrafted features. 



\section{Discussion \& Limitations}

In the developed system, since only single-channel EOG data is used, the desired performance cannot be achieved in multiple activity prediction. Classifiers have difficulty in distinguishing frowning, eyebrow raising and normal gaze activities. Given that these activities are related to the frontalis, corrugator supercilii, and procerus muscles, a single-channel cannot summarize all these muscle activities. However, the electrode around the frontalis muscle might additionally catch some unwanted neural activity signals. This can lead to an imprecise analysis of frontalis and orbicularis oculi muscle activities.

In an attempt to overcome the shortcomings of the device with a complex algorithmic design, it is concluded that at least a three-channel EOG structure should be investigated. This would provide insights into how to detect complex activity with a minimal sensing EOG device. One solution would be to place three electrodes on the face, encompassing a triangular area that spans the targeted muscles (frontalis and corrugator supercilii and procerus) and reference them to each other. By using three different electrodes to obtain information from all these muscles, a passive interaction based complex activity recognition could be performed, including mood estimation. Furthermore, as a contribution to future minimal sensing research, exploratory studies should be carried out. These would involve placing multiple electrodes on the face and exploring how different activities can be optimally estimated with the minimum number of electrodes, at which locations and with which cross-referencing strategies.

Furthermore, compared to classification-based EOG-HMI systems, a regression-based one that picks up even the slightest eye movements can provide much more rapid and accurate feedback. Since such EOG-based systems do not attempt to analyze cortical activity, it is simpler to develop remarkably faster systems. As an example P300 based systems require substantial amount of latencies (e.g. 300 ms in single-trial P300) \citep{b25}. Moreover, it necessitates an additional GUI to create a computer simulation for capturing the attention based event-related potentials (ERP). This, in turn, demands from the subjects a steady focus on the interface, apart from multiple-EEG electrodes on their head. Given that the primary aim is a controlling mouse cursor on the computer, reserving a portion of a screen for such GUI will not be preferable for usability. Although there are P300 based BCI systems which successfully determine ERPs and translate them into video game character movements \citep[]{b26,b27,b28}, interactions require GUI to detect voluntary ERP markers. This is precisely why patients, especially those who have motor disabilities, might require an alternative system independent of the GUI for rapid and convenient communication with the computer.


\printcredits

\section*{Declaration of competing interest}
The authors declare that they have no known competing financial interests or personal relationships that could have appeared to
influence the work reported in this paper.

\bibliographystyle{model1-num-names}

\bibliography{ourBibliography}

\end{document}